\documentclass[11pt]{article}

\usepackage{graphicx}
\usepackage{rotating}
\usepackage{amssymb}
\usepackage{amsmath}
\usepackage{mathptmx}
\usepackage[numbers]{natbib}
\usepackage{authblk}
\makeatletter

\bibpunct{}{}{,}{s}{}{,}

\begin{document}

\newcommand{\hdblarrow}{H\makebox[0.9ex][l]{$\downdownarrows$}-}
\title{A Spread-Spectrum SQUID Multiplexer}
\date{\vspace{-5ex}}
\author[1,2]{K.D. Irwin\thanks{irwin@stanford.edu}}
\author[1]{S. Chaudhuri}
\author[2]{H.-M. Cho}
\author[1]{C. Dawson}
\author[1]{S. Kuenstner}
\author[2]{D. Li}
\author[1]{C.J. Titus}
\author[1]{B.A. Young}

\affil[1]{Department of Physics, Stanford University, Stanford, CA 94305}
\affil[2]{SLAC National Accelerator Laboratory, Menlo Park, CA 94025}

\maketitle
\begin{abstract}
The Transition-Edge Sensors (TES) is a mature, high-resolution x-ray spectrometer technology that provides a much higher efficiency than dispersive spectrometers such as gratings and crystal spectrometers. As larger arrays are developed, time-division multiplexing schemes operating at MHz frequencies are being replaced by microwave SQUID multiplexers using frequency-division multiplexing at GHz frequencies. However, the multiplexing factor achievable with microwave SQUIDs is limited by the high slew rate on the leading edge of x-ray pulses. In this paper, we propose a new multiplexing scheme for high-slew-rate TES x-ray calorimeters: the spread-spectrum SQUID multiplexer, which has the potential to enable higher multiplexing factors, especially in applications with lower photon arrival rates.

\end{abstract}

\section{Introduction}

Transition Edge Sensors (TES) [\hspace{-4 pt} \citenum{irwin1995TES}] provide a unique combination of high spectral resolution and high efficiency for x-ray spectroscopy at light sources [\hspace{-4 pt} \citenum{ullom2015review}] and in x-ray astrophysics [\hspace{-4 pt} \citenum{porter2005x}]. TES systems are deployed at multiple x-ray light sources using time-division multiplexers (TDM)[\hspace{-4 pt} \citenum{chervenak1999superconducting}]. New generations of instruments require much higher multiplexing factors.

The spread-spectrum SQUID multiplexer (SSMux) can provide higher multiplexing factors in some high-slew-rate x-ray instruments by combining circuit elements developed for microwave SQUID multiplexers ($\mu$mux) [\hspace{-4 pt} \citenum{irwin2004microwave}, \hspace{-4 pt} \citenum{mates2017simultaneous}] and code-division SQUID multiplexers (CDM) [\hspace{-4 pt} \citenum{irwin2010code}, \hspace{-4 pt} \citenum{morgan2016code}]. The SSMux takes the signal from a TES and deliberately spreads it in the frequency domain to increase the achievable slew rate and multiplexing factor. 

We start by reviewing $\mu$mux ($\S$\ref{sec:umux}) and the limits it places on slew rate and MUX factors ($\S$\ref{sec:LineSpacing}) before describing the advantages of SSMux in high-slew-rate x-ray spectrometers ($\S$\ref{sec:SSMux}), and details of the implementation of systems based on SSMux, including a consideration of applications with higher count rates ($\S$\ref{sec:system}).

\section{Microwave SQUID multiplexers ($\mu$mux)} \label{sec:umux}
\begin{figure}[htbp]
\begin{center}
\includegraphics[width=0.8\linewidth, keepaspectratio]{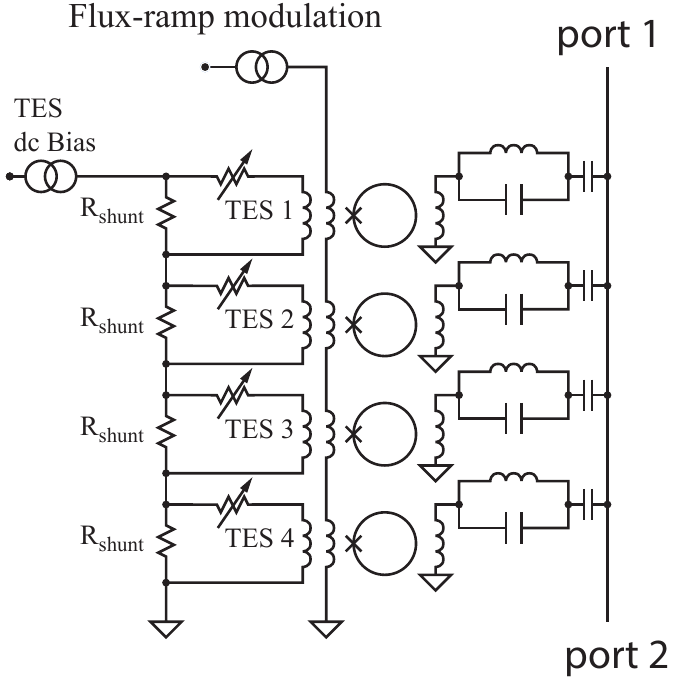}
\caption{The electrical schematic of a microwave SQUID multiplexer ($\mu$mux). A four-pixel implementation is shown in this example. A common dc TES bias current is applied on the left of the figure. The bias current passes through the parallel combination of the TES detectors and small shunt resistors $R_{\rm shunt}$, so that the TES detectors are voltage biased. The bias current passes in series through all detectors and their parallel shunts. The current flowing through each TES applies a flux to a dissipationless, non-hysteretic RF SQUID coupled to a microwave resonator. Each microwave resonator is tuned to a unique frequency. A comb of excitation frequencies tuned to each resonator is incident from ``port 1''. The transmitted signal, carrying the imprint of the status of each TES-coupled resonator, is carried out of ``port 2'' to the amplifier. A common sawtooth flux-ramp-modulation signal is applied to all SQUIDs.}
\end{center}
\label{fig1}
\end{figure}
Time-division multiplexing (TDM) schemes are used to read out TES arrays in deployed x-ray spectrometers [\hspace{-4 pt} \citenum{doriese2016developments}]. This approach, however, has limited scalability. The modest total bandwidth ($\sim 10$ MHz) limits the number of signals that can be multiplexed in one wire. In contrast, at microwave frequencies, compact microwave-filter elements can be used, and the large total bandwidth makes it possible to multiplex more signals in each wire. 

In $\mu$mux (Fig. 1), a SQUID is placed at every pixel in a high-Q resonant circuit with a unique resonance frequency [\hspace{-4 pt} \citenum{irwin2004microwave}, \hspace{-4 pt} \citenum{mates2017simultaneous}]. In this approach, large arrays of TES detectors are frequency-division multiplexed with a pair of coaxial cables.
The response of the microwave SQUIDs is linearized by applying a common flux ramp to all SQUIDs [\hspace{-4 pt} \citenum{Mates2012}]. The flux ramp is a sawtooth with an amplitude of an integer number $n_{\Phi_0}$ of flux quanta. The detector signal is measured as a change in the phase of the periodic SQUID response. This phase change is a linear function of the detector signal and can be tracked through many flux quanta. 

The number of pixels that can be multiplexed in one amplifier channel and one pair of coaxial cables is determined by the available bandwidth of the amplifier and room-temperature readout electronics, and the frequency spacing between resonators. For example, for 1 MHz resonator spacing and a typical bandwidth of 4--8 GHz, 4,000 resonator-coupled TESs could be read out in a pair of coaxial cables. Modern room-temperature RF electronics are able to synthesize and track this number of tones [\hspace{-4 pt} \citenum{Kernasovskiy2017}].

\section{Limitations on multiplexing factor from slew rate}\label{sec:LineSpacing}

The resonator spacing in $\mu$mux is generally much larger than the frequency content of the signals to be multiplexed, leading to low Shannon efficiency in the multiplexer circuit. The Shannon efficiency of a GHz multiplexer based on superconducting resonators, such as an MKID or microwave SQUID, is typically $\sim 10^{-5}$ [\hspace{-4 pt} \citenum{irwin2009shannon}]. Significant improvement is possible and desirable. The resonator spacing is typically limited by either fabrication nonuniformity in the frequency of the resonators, or by the bandwidth of the flux-ramp modulation in detector applications requiring high slew rate.

In many applications, including measurements of the Cosmic Microwave Background, TES bolometers are used to measure slowly varying signals. In these cases, the resonator line spacing is limited by fabrication nonuniformity. If the resonator spacing is too close, or the fabrication process too variable, random variation in resonator position can cause resonator line reordering and collision, decreasing array yield and causing difficulty in identifying which resonator couples to which pixel. Advances in fabrication techniques are improving the resonator line spacing, including the implementation of techniques for a final fabrication step to trim the resonator frequencies after cryogenic measurement. It should be practical to place resonators on $<1$~MHz spacing in the 4--8 GHz range in future arrays.

However, in applications requiring high slew rate, including x-ray spectroscopy, the resonator spacing can instead be limited by the bandwidth required to track signals with high slew rate. As described in section $\S$\ref{sec:umux}, the TES detector signal flux is added to a sawtooth ramp in flux in the input of the SQUID. The frequency of the resonator varies periodically with its input flux, with period equal to the magnetic flux quantum $\Phi_0$. As the flux ramps, the resonator sweeps through its frequency range. The TES signal is measured as a phase shift in this variation. If the TES signal flux varies by more than $\Phi_0$ in too short a period of time, the demodulation algorithm will lose the ability to deconvolve the flux-ramp from the input signal, leading to a flux-jump in the recorded detector signal. 

Conceptually, one ``sample'' of the input signal is computed for each repetition of the flux-ramp sawtooth through $n_{\Phi_0}$ flux quanta. If the flux from the TES signal changes by more than $\epsilon \Phi_0$ during this repetition period, the maximum ``error signal'' of the system is exceeded. Typically, $n_{\Phi_0}=2$ and $\epsilon \approx 0.5$. The maximum flux slew rate that can be tracked by a microwave SQUID is
\begin{equation}\label{eqn:fluxslew}
	\left. \frac{d\Phi}{dt}  \right\vert_{\rm max} = \left. M\frac{dI}{dt}  \right\vert_{\rm max} =\epsilon f_s \Phi_0,
\end{equation}
where $f_s$ is the flux-ramp sawtooth repetition frequency, and $M$ is the mutual inductance of the input coil coupling to the TES current $I$.

The bandwidth required for each TES pixel is
\begin{equation}\label{eqn:BWpix}
  BW_{\rm pix}=2 S f_s n_{\Phi_0},
\end{equation}
where the 2 arises because there are sidebands on both sides of the central frequency, and $S$ is the normalized spacing between resonators ($S \gtrsim 10$ to minimize crosstalk) (see [\hspace{-4 pt} \citenum{Mates2012}] for a detailed discussion of flux-ramp modulation).

Combining Eq. (1) and (2), we arrive at an equation for the required slew-rate-limited bandwidth per pixel in a TES array read out by a microwave SQUID:
\begin{equation}\label{eqn:BWpix2}
	BW_{\rm pix} = \frac{2 S n_{\Phi_0}}{\epsilon} 
	\frac{M}{\Phi_0}\left. \frac{dI}{dt}  \right\vert_{\rm max}.
\end{equation}

The number of pixels that can be read out in total bandwidth $BW_{\rm tot}$ is thus

\begin{align}\label{eqn:Npix}
N_{\rm pix}&= \frac{BW_{\rm tot}}{2 S n_{\Phi_0} \times f_s} \\
& = \frac{\epsilon \Phi_0 BW_{\rm tot}}{2 S n_{\Phi_0} \times \left. M\frac{dI}{dt}  \right\vert_{\rm max} }
\end{align}

As an example, consider the calorimeters that are envisioned for the LPA1 configuration of the Athena X-Ray Integral Field Unit (X-IFU) [\hspace{-4 pt} \citenum{Barret2016}], with designed maximum current slew rate on the pulse leading edge of $dI/dt = 0.4$ A/s. In this case, the required noise performance is achieved in a modern $\mu$mux with mutual inductance $M=230$pH. Taking $n_{\Phi_0}=2$, $S=10$, and $\epsilon=0.5$, Eq. (3) gives $BW/$pix $\sim 4$ MHz. From Eq. (5), in a bandwidth of 4--8 GHz, 1000 pixels could be multiplexed in each pair of coaxial cables, limited by the required slew rate on the pulse leading edge. While this is a good multiplexing factor, it is at least 4 times worse than achievable resonator frequency packing. Larger multiplexing factors are desirable. As we show in the next section, increases in the multiplexing factor can be enabled by the SSMux.

\section{Spread-spectrum multiplexer (SSMux)}\label{sec:SSMux}

As shown in $\S$\ref{sec:LineSpacing}, the bandwidth required by each pixel in a $\mu$mux circuit used in an x-ray spectrometer is determined by the maximum slew rate on the leading edge of the pulse. However, at any given time, few pixels are in the steep part of the leading edge of a pulse, where the slew rate is highest. The fraction of pixels on the high-slew-rate part of a pulse is especially small in photon-starved applications where the overall count rate is low (e.g. some x-ray astronomy missions). In this section we show that the MUX factor and slew-rate budget can be increased by spreading the flux signal from each pixel over multiple resonators in a Walsh code [\hspace{-4 pt} \citenum{walsh1923closed}], and calculate the advantage in the photon-starved limit. In $\S$\ref{sec:system}, we discuss the advantage that can be achieved in applications with higher photon rates.

\begin{figure}[htbp]
\begin{center}
\includegraphics[width=0.8\linewidth, keepaspectratio]{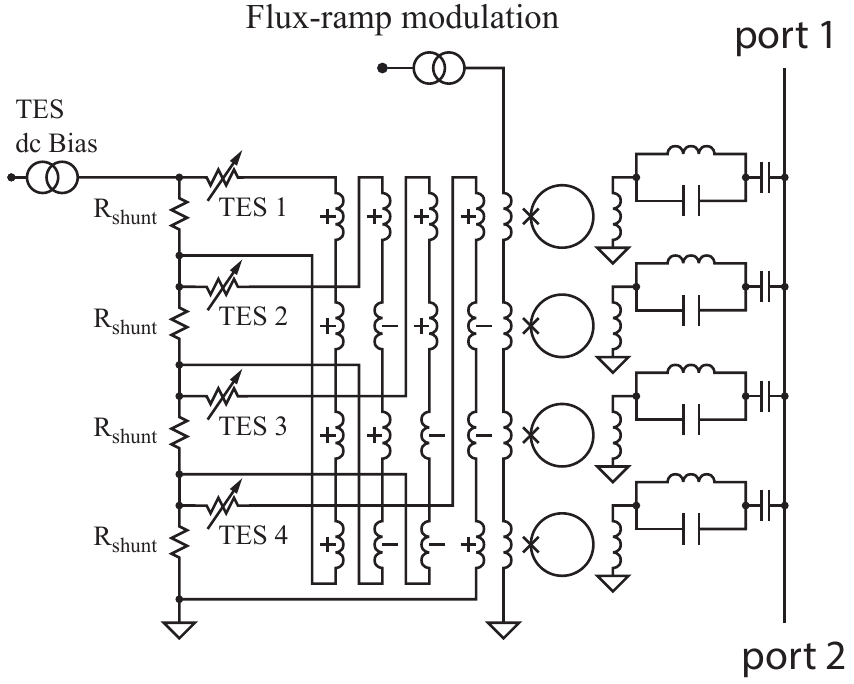}
\caption{The electrical schematic of a spread-spectrum SQUID multiplexer (SSMux). A four-pixel implementation is shown in this example. The detector bias, flux-ramp modulation, and microwave SQUID readout operate the same way as the simple $\mu$mux circuit shown in Fig. 1. In the SSMux, however, the current from each TES is incident on all four SQUIDs shown, with coupling polarities modulating in a Walsh code. There is one SQUID for each TES, as in $\mu$mux, but the flux-slew burden from the leading edge of an x-ray pulses is shared among all SQUIDs.
}
\end{center}
\label{fig2}
\end{figure}

In the SSMux, the signal from each TES is coupled to $N_{\rm ss}$ resonators (see Fig. 2). At the same time, each individual SQUID resonator is coupled to $N_{\rm ss}$ different TESs in a Walsh code. In this scheme, the total number of microwave SQUIDs is still equal to the number of TES detectors, but the high slew rate of a pixel on the steep rising edge of a pulse is divided between $N_{\rm ss}$ different resonators, reducing the slew rate required in each, and thus reducing the bandwidth each resonator requires. The reduction in required bandwidth makes it possible to place the resonators closer, allowing a higher MUX factor in each pair of coaxial cables.

Walsh code-division multiplexing of TES detectors into time-division multiplexed SQUIDs is now well established [\hspace{-4 pt} \citenum{irwin2010code}, \hspace{-4 pt} \citenum{morgan2016code}]. The pattern with which the $N_{\rm ss}$ detectors are coupled into the $N_{\rm ss}$ SQUIDs is an orthogonal Walsh code, so the combination can be inverted, extracting independent measurements of each TES. During each flux-ramp period, $N_{\rm ss}$ different measurements of each TES signal are made, with uncorrelated SQUID noise. When the Walsh code is inverted, combining the $N_{\rm ss}$ measurements reduces the effective SQUID noise amplitude by $\sqrt{N_{\rm ss}}$. Thus, the same overall signal-to-noise ratio can be achieved as in $\mu$mux with lower coupling to the current from each individual SQUID. The mutual inductance $M$ can be reduced as much as $M/\sqrt{N_{\rm ss}}$. As long as only one of the TESs in the Walsh set is in a high-slew-rate condition at a given time, the maximum flux slew rate applied to this resonator is reduced by $\sqrt{N_{\rm ss}}$. 

In CDM, unlike TDM, the SQUID will experience pulses with both increasing and decreasing flux (the two polarities in the Walsh code), so CDM must be biased in the middle of the SQUID response curve rather than near one extreme. Thus, in CDM, the maximum flux slew rate that can be tolerated without losing lock is degraded by approximately $\times 2$ relative to TDM. While SSMux also implements Walsh codes, it does not share this $\times 2$ slew-rate degradation. The maximum flux slew rate that can be tolerated without losing lock is the same as microwave SQUIDs (Eqn. \ref{eqn:fluxslew}) since it is flux-ramp modulated rather than biased at a fixed input flux.

Because of the details of the modulation and filter functions, the effective SQUID noise referred to the input is modestly degraded in TDM, CDM, microwave SQUID, and SSMux circuits. In TDM and CDM temporally switched multiplexers, because a one-pole $L/R$ filter is used rather than the ideal temporal boxcar filter, the noise amplitude in these circuits is increased by $\sqrt{\pi}$ [\hspace{-4 pt} \citenum{doriese2006progress}] . Because of the signal-to-noise inefficiencies of approximately sinusoidal frequency modulation, the input-referred noise amplitude in $\mu$mux is degraded by $\sim \sqrt{2} $[\hspace{-4 pt} \citenum{Mates2012}]. SSMux is read out with flux-ramp modulation but is not switched, so SSMux shares the $\sim \sqrt{2}$ noise degradation with microwave SQUIDs, but not the $\sqrt{\pi}$ degradation of TDM and CDM, as it is not temporally switched.

Taking all of these factors into account, the bandwidth of each resonator can be reduced by up to $\sqrt{N_{\rm ss}}$ relative to $\mu$mux. The bandwidth required per pixel can be as low as
\begin{equation}
	BW_{\rm pix}(N_{\rm ss}) \gtrapprox \frac{2 S n_{\Phi_0}}{\epsilon} 
	\frac{M}{\Phi_0} \frac{1}{\sqrt{N_{\rm ss}}}  \left. \frac{dI}{dt}  \right\vert_{\rm max},
\end{equation}
and the maximum number of pixels that can be multiplexed in a pair of coaxial cables can be as high as
\begin{equation}
N_{\rm pix}(N_{\rm ss}) \lessapprox \frac{\epsilon \Phi_0  BW_{\rm tot} \sqrt{N_{\rm ss}}} {2 S n_{\Phi_0} \times \left. M\frac{dI}{dt}  \right\vert_{\rm max} }.
\end{equation}

In the previous section, we calculated that in the example of the LPA1 configuration of the Athena X-IFU, the bandwidth per pixel required by slew rate is $BW_{\rm pix} \approx 4$ MHz. The implementation of a SSMux with $N_{\rm ss}=16$ would reduce the required bandwidth by a factor of $\sqrt{N_{\rm ss}}=4$ to $BW_{\rm pix} \approx 1$ MHz, and increase the multiplex factor to approximately 4,000 per coaxial cable pair.

\section{System implementation}\label{sec:system}

The increased requirement on the room-temperature electronics for SSMux relative to $\mu$mux is modest. For each coded group of $N_{\rm ss}$ pixels, an additional $N_{\rm ss} \times N_{\rm ss}$ multiplies is required for demultiplexing the group. Thus, SSMux requires an additional $N_{\rm ss}$ multiplies per pixel. These computational requirements are likely to be subdominant to the flux-ramp demultiplexing.

The feedlines in both $\mu$mux and SSMux carry signal tones at each resonator frequency. The nonlinearity of the follow-on amplifier creates intermodulation products in the signal band that can degrade signal to noise. For the same number of pixels, the challenge of mitigating intermodulation products is the same in SSMux as in $\mu$mux with all else held fixed, as the same microwave excitation powers are used in each case, and the number of tones is the same. However, in SSMux, the number of pixels multiplexed on each feedline may be increased by as much as $\sqrt{N_{\rm ss}}$. The number of third-order intermodulation products increases as the cube of the number of resonators on the feedline, increasing the challenge of mitigating intermodulation products. However, this challenge is manageable.


As calculated in $\S$\ref{sec:SSMux}, SSMux has clear advantages in the limit of photon-starved applications. It can also be useful in applications at higher photon-arrival rates. Optimizing such a design requires detailed analysis of source models and resource requirements for different parts of the system. As an example of this optimization, we present a very simple case: a detector array for a free-electron laser (FEL) such as the Linac Coherent Light Source, in which the photons arrive at essentially the same time, only pixels that receive exactly one photon in a given repetition provide useful data, and the pixels recover before the next photon repetition. We further assume for this simplified analysis that the system capabilities are limited only by the available bandwidth of the readout electronics and coaxial cabling, so that SSMux allows larger arrays to be instrumented in the same bandwidth.

The number of photons received by each pixel in this case is determined by a Poisson distribution. The probability of a pixel receiving zero photons in one repetition is thus $P(0)=e^{-\lambda}$, and the probability of one photon is $P(1)=\lambda e^{-\lambda}$, where $\lambda$ is the average number of photons per repetition in each pixel. For conventional $\mu$mux ($N_{\rm ss}=1$), $N_{\rm pix}(1)$ pixels can be accomodated in bandwidth $BW_{\rm tot}$, so the total number of useful counts per repetition $C_{\rm rep}(1)$ for $N_{\rm ss}=1$ in this bandwidth is 

\begin{equation}\label{eqn:FELnorm}
C_{\rm rep}(1) = N_{\rm pix}(1) \lambda e^{-\lambda}.
\end{equation}

As described above, if SSMux is used ($N_{\rm ss}>1$), the number of pixels that can occupy the same bandwidth $BW_{\rm tot}$ at the same slew rate is $N_{\rm pix}(N_{\rm ss})=N_{\rm pix}(1) \sqrt{N_{\rm ss}}$. The total number of useful counts per repetition in the same bandwidth as the readout in Eqn. \ref{eqn:FELnorm} can then be calculated. The probability that each Walsh-coded group of $N_{\rm ss}$ pixels will receive one photon in one pixel, and zero photons in the others is multiplied by the number of such coded groups, increasing the total number of useful counts $C_{\rm rep}$ by the factor 


\begin{equation}\label{eqn:FELssmux2}
\frac{C_{\rm rep}(N_{\rm ss})}{C_{\rm rep}(1)} = \sqrt{N_{\rm ss}} e^{(1-N_{\rm ss})\lambda}.
\end{equation}

The number of useful counts in Eqn. \ref{eqn:FELssmux2} is maximized for 

\begin{equation}\label{eqn:Nopt}
N_{\rm ss\_opt}=1/(2 \lambda),
\end{equation}
for vales of $\lambda$ where $N_{\rm ss\_opt}$ is an integer. 
Thus, even for average number of photons per repetition as high as $\lambda = 0.25$, for which $N_{\rm ss\_opt}=2$,
the total useful count rate in this simplified model for an FEL array can be increased by the use of SSMux rather than $\mu$mux with the same readout bandwidth.

\section{Conclusion}

The spread-spectrum SQUID multiplexer shares the flux-slew burden from the leading edge of an x-ray pulse across multiple SQUID resonators at different frequencies. By spreading the signal to a wider frequency range, the SSMux can enable higher slew rates and/or higher MUX factors. In photon-starved conditions, the full factor of $\sqrt{N_{\rm ss}}$ improvement in multiplexing factor is achieved. The SSMux may also improve performance at higher photon-arrival rates. The SSMux can be combined with hybrid multiplexing schemes, such as TDMA hybrid multiplexers [\hspace{-4 pt} \citenum{reintsema2008tdma}] or CDMA hybrid multiplexers [\hspace{-4 pt} \citenum{irwin2012advanced}], which multiplex multiple TES detectors in each SQUID resonator to increase their slew-rate handling capability.

\section{Acknowledgements}
This work was supported in part by the DOE Office of Basic Energy Sciences Scientific User Facilities Division Accelerator and Detector R\&D program, and by NASA under grant numbers NNX15AT02G and NNX16AH89G.

\pagebreak


\begin{thebibliography}{99}



\bibitem{irwin1995TES}
K.D. Irwin,  {\it Appl. Phys. Lett.}  \textbf{66}, 1998, (1995), DOI:10.1063/1.113674.

\bibitem{ullom2015review}
J.N. Ullom, and D.A. Bennett, {\it Superc. Sci. and Tech.} \textbf{28}, 084003, (2015), DOI:10.1088/0953-2048/28/8/084003.

\bibitem{porter2005x}
F.S. Porter, G.V. Brown, and J. Cottam, in {\it Cryogenic Particle Detection}, Springer Topics in Applied Physics, \textbf{99} 359, (2005), DOI:10.1007/10933596\textunderscore8.


\bibitem{chervenak1999superconducting}
J.A. Chervenak, K.D. Irwin, E.N. Grossman, J.M. Martinis, C.D. Reintsema, and M.E. Huber, {\it Appl. Phys. Lett.} \textbf{74}, 4043, (1999), DOI:10.1063/1.123255.

\bibitem{irwin2004microwave}
K.D. Irwin and K.W. Lehnert,  {\it Appl. Phys. Lett.}  \textbf{85}, 2107, (2004), DOI:10.1063/1.1791733.

\bibitem{mates2017simultaneous}
J.A.B. Mates, D.T. Becker, D.A. Bennett, B.J. Dober, J.D. Gard, J.P. Hays-Wehle, J.W. Fowler, G.C. Hilton, C.D. Reintsema, D.R. Schmidt, D.S. Swetz, L.R. Vale, and J.N. Ullom,  {\it Appl. Phys. Lett.}  \textbf{111}, 062601, (2017), DOI:10.1063/1.4986222.

\bibitem{irwin2010code}
K.D. Irwin, M.D. Niemack, J. Beyer, H.~M. Cho, W.B. Doriese, G.C. Hilton, C.D. Reintsema, D.R. Schmidt, J.N. Ullom, and L.R. Vale,  {\it Superc. Sci. and Tech.}  \textbf{23}, 034004, (2010), DOI:10.1088/0953-2048/23/3/034004.

\bibitem{morgan2016code}
K.M. Morgan, B.K. Alpert, D.A. Bennett, E.V. Denison, W.B. Doriese, J.W. Fowler, J.D. Gard, G.C. Hilton, K.D. Irwin, Y.I. Joe, G.C. O'Neil, C.D. Reintsema, D.R. Schmidt, J.N. Ullom, and D.S. Swetz, {\it Appl. Phys. Lett.}  \textbf{109}, 112604, (2016), DOI:10.1063/1.4962636.

\bibitem{doriese2016developments}
W. B. Doriese, K. M. Morgan, D. A. Bennett, E. V. Denison, C. P. Fitzgerald, J. W. Fowler, J. D. Gard,
J. P. Hays-Wehle, G. C. Hilton, K. D. Irwin, Y. I. Joe, J. A B Mates, G. C. O’Neil, C. D. Reintsema,
N. O. Robbins, D. R. Schmidt, D. S. Swetz, H. Tatsuno, L. R. Vale, and J. N. Ullom,  {\it J. Low Temp. Phys.}  \textbf{184}, 389, (2016), DOI:10.1007/s10909-015-1373-z.

\bibitem{Mates2012}
J.A.B Mates, K. D. Irwin, L. R. Vale, G. C. Hilton, J. Gao, and K. W. Lehnert, {\it J. Low Temp. Phys.} \textbf{167}, 707, (2012), DOI:10.1007/s10909-012-0518-6.

\bibitem{Kernasovskiy2017}
S.A. Kernasovskiy, S. Kuenstner, E. Karpel,
Z. Ahmed, D.D. Van Winkle, S. Smith, J.
Dusatko, J.C. Frisch, S. Chaudhuri, H. M.
Cho, B. Dober, S.W. Henderson, G. Hilton, J.
Hubmayr, K. D. Irwin, C. L. Kuo, D. Li, J. A.
B. Mates, M. Nasr, S. Tantawi, J. Ullom, L.
Vale, and B. A. Young, {\it J. Low Temp.
Phys.} This Special Issue (2017).

\bibitem{irwin2009shannon}
K.D. Irwin, {\it AIP Conf. Proc.} \textbf{1185}, 229, (2009), DOI:10.1063/1.3292320.

\bibitem{Barret2016}
D. Barret \emph{et al.}, {\it Proc. SPIE. 9905} \textbf{9905}, 99052F, (2016), DOI:10.1117/12.2232432.

\bibitem{walsh1923closed}
J.L. Walsh,  {\it Amer. Journ. Math.}  \textbf{45}, 5, (1923), DOI:10.2307/2387224.

\bibitem{doriese2006progress}
W.B. Doriese, J.A. Beall, W.D. Duncan, L. Ferreira, G.C. Hilton, K.D. Irwin, C.D. Reintsema, J. Ullom, L. Vale. and Y. Xu, {\it Nucl. Instr. and Meth.} \textbf{A559}, 808, (2006), DOI:10.1016/j.nima.2005.12.146.

\bibitem{reintsema2008tdma}
C.D. Reintsema, J. Beall, W.B. Doriese, W. Duncan, L. Ferreira, G.C. Hilton,
K.D. Irwin, D. Schmidt, J. Ullom, L. Vale. and Y. Xu, {\it J. Low Temp. Phys.} \textbf{151}, 927, (2008), DOI:10.1007/s10909-008-9769-7.

\bibitem{irwin2012advanced}
K. D. Irwin, H. M. Cho, W. B. Doriese, J. W. Fowler, G. C. Hilton, M. D.
Niemack, C. D. Reintsema, D. R. Schmidt, J. N. Ullom, and L. R. Vale, {\it J. Low Temp. Phys.} \textbf{167}, 588, (2012), DOI:10.1007/s10909-012-0586-7.






\end{thebibliography}
\end{document}